\documentclass{emulateapj}
\usepackage{apjfonts}
\usepackage{rotating}

\lefthead{CHOI ET AL.} 
\righthead{Planet/Binary Degeneracy}

\begin{document}
\title{A NEW TYPE OF AMBIGUITY IN THE PLANET AND BINARY 
INTERPRETATIONS OF CENTRAL PERTURBATIONS OF HIGH-MAGNIFICATION 
GRAVITATIONAL MICROLENSING EVENTS}

\author{
J.-Y. Choi$^{1}$,
I.-G. Shin$^{1}$,
C. Han$^{1,G1,S}$,
A. Udalski$^{O01,G1}$, 
T. Sumi$^{M01,G2}$,        
A. Gould$^{u01,G3}$,
V. Bozza $^{702,722,G4}$,
M. Dominik $^{703,\star,G4}$,
P. Fouqu\'e$^{P01,G5}$,
K. Horne$^{703,G6}$,\\ 
and\\
M.\,K. Szyma{\'n}ski$^{O01}$,
M. Kubiak$^{O01}$, 
I. Soszy{\'n}ski$^{O01}$,
G. Pietrzy{\'n}ski$^{O01,O02}$, 
R. Poleski$^{O01}$, 
K. Ulaczyk$^{O01}$, 
P. Pietrukowicz$^{O01}$,
S. Koz{\l}owski$^{O01}$,
J. Skowron$^{u01}$,
{\L}. Wyrzykowski$^{O01,O03}$\\ 
(The OGLE Collaboration),\\
F. Abe$^{M02}$,
D.P. Bennett$^{M03}$,
I.A. Bond$^{M04}$,
C.S. Botzler$^{M05}$,
P. Chote$^{M06}$,
M. Freeman$^{M05}$,
A. Fukui$^{M07}$,
K. Furusawa$^{M02}$,
Y. Itow$^{M02}$,
S. Kobara$^{M02}$,
C.H. Ling$^{M04}$
K. Masuda$^{M02}$,
Y. Matsubara$^{M02}$,
N. Miyake$^{M02}$,
Y. Muraki$^{M02}$,
K. Ohmori$^{M02}$,
K. Ohnishi$^{M08}$,
N.J. Rattenbury$^{M05}$,
To. Saito$^{M10}$,
D.J. Sullivan$^{M06}$,
D. Suzuki$^{M01}$,
K. Suzuki$^{M02}$,
W.L. Sweatman$^{M04}$,
S. Takino$^{M02}$,
P.J. Tristram$^{M06}$,
K. Wada$^{M01}$,
P.C.M. Yock$^{M05}$\\
(The MOA Collaboration),\\
D.M. Bramich$^{R04}$,
C. Snodgrass$^{R06}$,
I.A. Steele$^{R05}$, 
R.A. Street$^{R02}$, 
Y. Tsapras$^{R02,R03}$\\ 
(The RoboNet Collaboration),\\
K.A. Alsubai$^{701}$,
P. Browne$^{703}$,
M.J. Burgdorf$^{704,705}$,
S. Calchi Novati$^{702, 706}$,
P. Dodds$^{703}$,
S. Dreizler$^{707}$,
X.-S. Fang$^{708}$,
F. Grundahl$^{709}$,
C.-H. Gu$^{708}$,
S. Hardis$^{710}$,
K. Harps{\o}e $^{710,711}$,
T.C. Hinse$^{710,712,713}$,
A. Hornstrup$^{714}$,
M. Hundertmark$^{703,707}$,
J. Jessen-Hansen$^{709}$,
U.G. J{\o}rgensen$^{710,711}$,
N. Kains$^{715}$,
E. Kerins$^{716}$,
C. Liebig$^{703}$,
M. Lund$^{709}$,
M. Lunkkvist$^{709}$,
L. Mancini$^{717,718}$,
M. Mathiasen$^{710}$,
M.T. Penny$^{716, u01}$,
S. Rahvar$^{719,728}$,
D. Ricci$^{720}$,
G. Scarpetta$^{702,706,722}$,
J. Skottfelt$^{710}$,
J. Southworth$^{725}$,
J. Surdej$^{720}$,
J. Tregloan-Reed$^{725}$,
J. Wambsganss$^{726}$,
O. Wertz$^{720}$\\
(The MiNDSTEp Consortium),\\
L. A. Almeida$^{u02}$,
V. Batista$^{u01}$,
G. Christie$^{u03}$,
D.L. DePoy$^{u04}$,
Subo Dong$^{u05}$,
B.S. Gaudi$^{u01}$,
C. Henderson$^{u01}$,
F. Jablonski$^{u02}$,
C.-U. Lee$^{713}$, 
J. McCormick$^{u07}$,
D. McGregor$^{u01}$,
D. Moorhouse$^{u08}$,
T. Natusch$^{u03,u09}$,
H. Ngan$^{u03}$,
S.-Y. Park$^{1}$,
R.W. Pogge$^{u01}$,
T.-G. Tan$^{u10}$,
G. Thornley$^{u08}$,
J.C. Yee$^{u01}$\\
(The $\mu$FUN Collaboration),\\
M.D. Albrow$^{P04}$,
E. Bachelet$^{P01}$,
J.-P. Beaulieu$^{P03}$,
S. Brillant$^{P08}$,
A. Cassan$^{P03}$,
A.A. Cole$^{P05}$,
E. Corrales$^{P03}$,
C. Coutures$^{P03}$,
S. Dieters$^{P05}$,
D. Dominis Prester$^{P09}$,
J. Donatowicz$^{P10}$,
J. Greenhill$^{P05}$,
D. Kubas$^{P08,P03}$,
J.-B. Marquette$^{P03}$,
J.W. Menzies$^{P02}$,
K.C. Sahu$^{P06}$,
M. Zub$^{726}$\\
(The PLANET Collaboration),\\
}

\bigskip\bigskip
\affil{$^{1}$Department of Physics, Institute for Astrophysics, Chungbuk National University, Cheongju 371-763, Korea}
\affil{$^{O01}$Warsaw University Observatory, Al. Ujazdowskie 4, 00-478 Warszawa, Poland}
\affil{$^{O02}$Universidad de Concepci\'{o}n, Departamento de Astronomia, Casilla 160--C, Concepci\'{o}n, Chile}
\affil{$^{O03}$Institute of Astronomy, University of Cambridge, Madingley Road, Cambridge CB3 0HA, United Kingdom}
\affil{$^{M01}$Department of Earth and Space Science, Osaka University, Osaka 560-0043, Japan}
\affil{$^{M02}$Solar-Terrestrial Environment Laboratory, Nagoya University, Nagoya, 464-8601, Japan} 
\affil{$^{M03}$University of Notre Dame, Department of Physics, 225 Nieuwland Science Hall, Notre Dame, IN 46556-5670, USA} 
\affil{$^{M04}$Institute of Information and Mathematical Sciences, Massey University, Private Bag 102-904, North Shore Mail Centre, Auckland, New Zealand}
\affil{$^{M05}$Department of Physics, University of Auckland, Private Bag 92-019, Auckland 1001, New Zealand} 
\affil{$^{M06}$School of Chemical and Physical Sciences, Victoria University of Wellington, PO BOX 60, Wellington, New Zealand} 
\affil{$^{M07}$Okayama Astrophysical Observatory, NAOJ, Okayama 719-0232, Japan} 
\affil{$^{M08}$Nagano National College of Technology, Nagano 381-8550, Japan}
\affil{$^{M10}$Tokyo Metropolitan College of Aeronautics, Tokyo 116-8523, Japan} 
\affil{$^{R02}$Las Cumbres Observatory Global Telescope Network, 6740B Cortona Dr, Goleta, CA 93117, USA}
\affil{$^{R03}$School of Physics and Astronomy, Queen Mary University of London, Mile End Road, London, E1 4NS}
\affil{$^{R04}$European Southern Observatory, Karl-Schwarzschild-Str. 2, 85748 Garching bei M\"{u}nchen, Germany}
\affil{$^{R05}$Astrophysics Research Institute, Liverpool John Moores University, Liverpool CH41 1LD, UK}
\affil{$^{R06}$Max Planck Institute for Solar System Research, Max-Planck-Str. 2, 37191 Katlenburg-Lindau, Germany}
\affil{$^{701}$Qatar Foundation, P.O. Box 5825, Doha, Qatar}
\affil{$^{702}$Universit\`{a} degli Studi di Salerno, Dipartimento di Fisica "E.R. Caianiello", Via Ponte Don Melillo, 84084 Fisciano (SA), Italy}
\affil{$^{703}$SUPA, University of St Andrews, School of Physics \& Astronomy, North Haugh, St Andrews, KY16 9SS, United Kingdom}
\affil{$^{704}$Deutsches SOFIA Institut, Universit\"{a}t Stuttgart, Pfaffenwaldring 31, 70569 Stuttgart, Germany}
\affil{$^{705}$SOFIA Science Center, NASA Ames Research Center, Mail Stop N211-3, Moffett Field CA 94035, United States of America}
\affil{$^{706}$Istituto Internazionale per gli Alti Studi Scientifici (IIASS), Vietri Sul Mare (SA), Italy}
\affil{$^{707}$Institut f\"{u}r Astrophysik, Georg-August-Universit\"{a}t, Friedrich-Hund-Platz 1, 37077 G\"{o}ttingen, Germany}
\affil{$^{708}$National Astronomical Observatories/Yunnan Observatory, Joint laboratory for Optical Astronomy, Chinese Academy of Sciences, Kunming 650011, People's Republic of China}
\affil{$^{709}$Department of Physics and Astronomy, Aarhus University, Ny Munkegade 120, 8000 {\AA}rhus C, Denmark}
\affil{$^{710}$Niels Bohr Institute, University of Copenhagen, Juliane Maries vej 30, 2100 Copenhagen, Denmark}
\affil{$^{711}$Centre for Star and Planet Formation, Geological Museum, {\O}ster Voldgade 5, 1350 Copenhagen, Denmark}
\affil{$^{712}$Armagh Observatory, College Hill, Armagh, BT61 9DG, Northern Ireland, United Kingdom}
\affil{$^{713}$Korea Astronomy and Space Science Institute, 776 Daedukdae-ro, Yuseong-gu, Daejeon 305-348, Republic of Korea}
\affil{$^{714}$Danmarks Tekniske Universitet, Institut for Rumforskning og -teknologi, Juliane Maries Vej 30, 2100 K{\o}benhavn, Denmark}
\affil{$^{715}$ESO Headquarters, Karl-Schwarzschild-Str. 2, 85748 Garching bei M\"{u}nchen, Germany}
\affil{$^{716}$Jodrell Bank Centre for Astrophysics, University of Manchester, Oxford Road,Manchester, M13 9PL, UK}
\affil{$^{717}$Max Planck Institute for Astronomy, K\"{A}onigstuhl 17, 69117 Heidelberg, Germany}
\affil{$^{718}$International Institute for Advanced Scientific Studies, Vietri sul Mare (SA), Italy}
\affil{$^{719}$Department of Physics, Sharif University of Technology, P.~O.\ Box 11155--9161, Tehran, Iran}
\affil{$^{720}$Institut d'Astrophysique et de G\'{e}ophysique, All\'{e}e du 6 Ao\^{u}t 17, Sart Tilman, B\^{a}t.\ B5c, 4000 Li\`{e}ge, Belgium}
\affil{$^{722}$INFN, Gruppo Collegato di Salerno, Sezione di Napoli, Italy}
\affil{$^{725}$Astrophysics Group, Keele University, Staffordshire, ST5 5BG, United Kingdom}
\affil{$^{726}$Astronomisches Rechen-Institut, Zentrum f\"{u}r Astronomie der Universit\"{a}t Heidelberg (ZAH), M\"{o}nchhofstr.\ 12-14, 69120 Heidelberg, Germany}
\affil{$^{728}$Perimeter Institue for Theoretical Physics, 31 Caroline St. N., Waterloo, ON, N2L2Y5, Canada}
\affil{$^{\star}$Royal Society University Research Fellow}
\affil{$^{u01}$Department of Astronomy, Ohio State University, 140 West 18th Avenue, Columbus, OH 43210, United States of America}
\affil{$^{u02}$Instituto Nacional de Pesquisas Espaciais, S\~{a}o Jos\'{e} dos Campos, SP, Brazil}
\affil{$^{u03}$Auckland Observatory, Auckland, New Zealand}
\affil{$^{u04}$Dept.\ of Physics, Texas A\&M University, College Station, TX, USA}
\affil{$^{u05}$Institute for Advanced Study, Einstein Drive, Princeton, NJ 08540, USA}
\affil{$^{u07}$Farm Cove Observatory, Centre for Backyard Astrophysics, Pakuranga, Auckland, New Zealand}
\affil{$^{u08}$Kumeu Observatory, Kumeu, New Zealand}
\affil{$^{u09}$AUT University, Auckland, New Zealand}
\affil{$^{u10}$Perth Exoplanet Survey Telescope, Perth, Australia}
\affil{$^{P01}$IRAP, Universit\'e de Toulouse, CNRS, 14 Avenue Edouard Belin, 31400 Toulouse, France}
\affil{$^{P02}$South African Astronomical Observatory, P.O. Box 9 Observatory 7925, South Africa}
\affil{$^{P03}$UPMC-CNRS, UMR 7095, Institut d'Astrophysique de Paris, 98bis boulevard Arago, F-75014 Paris, France}
\affil{$^{P04}$University of Canterbury, Department of Physics and Astronomy, Private Bag 4800, Christchurch 8020, New Zealand}
\affil{$^{P05}$University of Tasmania, School of Mathematics and Physics, Private Bag 37, Hobart, TAS 7001, Australia}
\affil{$^{P06}$Space Telescope Science Institute, 3700 San Martin Drive, Baltimore, MD 21218, United States of America}
\affil{$^{P07}$Department of Physics, Massachussets Institute of Technology, 77 Mass. Ave., Cambridge, MA 02139, USA}
\affil{$^{P08}$European Southern Observatory, Casilla 19001, Vitacura 19, Santiago, Chile}
\affil{$^{P09}$Department of Physics, University of Rijeka, Omladinska 14, 51000 Rijeka, Croatia}
\affil{$^{P10}$Technische Universit\"{a}t Wien, Wieder Hauptst. 8-10, A-1040 Vienna, Austria}
\affil{${G1}$:The OGLE Collaboration.}
\affil{${G2}$:The MOA Collaboration.}
\affil{${G3}$:The $\mu$Fun Collaboration.}
\affil{${G4}$:The MiNDSTEp Consortium.}
\affil{${G5}$:The PLANET Collaboration.}
\affil{${G6}$:The RoboNet Collaboration.}
\affil{$^{S}$Author to whom any correspondence sholud be addressed.}

\begin{abstract}
High-magnification microlensing events provide an important channel 
to detect planets. Perturbations near the peak of a high-magnification 
event can be produced either by a planet or a binary companion. It is known 
that central perturbations induced by both types of companions can be 
generally distinguished due to the basically different magnification 
pattern around caustics. In this paper, we present a case of central 
perturbations for which it is difficult to distinguish the planetary 
and binary interpretations. The peak of a lensing light curve  
affected by this perturbation appears to be blunt and flat. For a 
planetary case, this perturbation occurs when the source trajectory 
passes the negative perturbation region behind the back end of an 
arrowhead-shaped central caustic. For a binary case, a similar perturbation 
occurs for a source trajectory passing through the negative perturbation 
region between two cusps of an astroid-shaped caustic. We demonstrate 
the degeneracy for 2 high-magnification events of OGLE-2011-BLG-0526 
and OGLE-2011-BLG-0950/MOA-2011-BLG-336. For OGLE-2011-BLG-0526, the 
$\chi^2$ difference between the planetary and binary model is $\sim$ 3, 
implying that the degeneracy is very severe. For OGLE-2011-BLG-0950/MOA-2011-BLG-336, 
the stellar binary model is formally excluded with $\Delta \chi^2 \sim$ 105 
and the planetary model is preferred. However, it is difficult to claim 
a planet discovery because systematic residuals of data from 
the planetary model are larger than the difference between the planetary and 
binary models. Considering that 2 events observed during a single season suffer from 
such a degeneracy, it is expected that central perturbations experiencing 
this type of degeneracy is common.
\end{abstract}


\section{Introduction}
\label{sec:two}

Microlensing constitutes one of the major methods to detect and characterize 
extrasolar planets \citep{mao91, gould92a}. The method is sensitive to 
planets that are difficult to be detected by using other methods such 
as cool planets at or beyond the snow line \citep{bond04, gaudi08, dong09, 
sumi10, muraki11} and planets at large distances \citep{janczak10}. 
It is also sensitive to low-mass planets \citep{beaulieu06, bennett08}, 
making it possible to detect terrestrial planets from ground observations. 
Due to the weak dependence on the host-star brightness, it also enables to 
detect planets around low-mass stars down to M-type dwarfs \citep{udalski05, 
miyake11, batista11} and even to sub-stellar mass objects. In addition, 
it is the only method that can detect old planetary-mass objects that are 
not bound to stars \citep{sumi11}. Therefore, microlensing is important 
for the complete census of the frequency and properties of planets \citep{gould10, cassan12}.

Current microlensing planet searches are being conducted based on a specially 
designed strategy where survey and follow-up observations work in close coordination. 
There are two main reasons for this strategy. The first reason is that the probability 
of a lensing event is very low. For a star located in the Galactic bulge, toward 
which planetary microlensing searches are being conducted, the chance to detect a lensed 
star at a specific time is of order $10^{-6}$ \citep{udalski94, alcock00}. Considering 
that a planet can be detected for a small fraction of lensing events, it is essential 
to maximize the detection rate of lensing events to increase the rate of planet detections. 
Survey observations are designed for this purpose by monitoring a large area of 
the Galactic bulge field. The second reason for the survey/follow-up strategy is 
that the duration of a planetary signal is short. The planetary signal is a short-term 
perturbation to the smooth standard light curve of the primary-induced lensing event. 
To densely cover planetary perturbations, follow-up observations are designed 
to focus on events detected by survey observations. 

Under the current strategy of microlensing searches, high-magnification events are 
important targets for follow-up observations. A typical number of events alerted at 
a certain time by survey experiments is of order 10. Considering that each event 
typically lasts for several dozens of days, it is difficult to follow all alerted 
events with a restricted number of telescopes. To maximize the planet detection 
efficiency, therefore, priority is given to events for which the planet detection 
probability is high. Currently, the highest priority is given to high-magnification 
events. For a lens with a planet, there exist two sets of disconnected caustics, 
where one set is located away from the planet-host star (planetary caustic) while 
the other set is always located close to the host star (central caustic). For 
a high-magnification event, the sensitivity to a planetary companion is very high 
because the source trajectory always passes close to the perturbation region around 
the central caustic induced by the planet \citep{griest98}. The efficiency of the 
strategy focusing on high-magnification events is demonstrated by the fact that 
7 out of 13 microlensing planets detected as of the end of 2011 were detected through 
this channel. 

\begin{figure}[t]
\epsscale{1.15}
\plotone{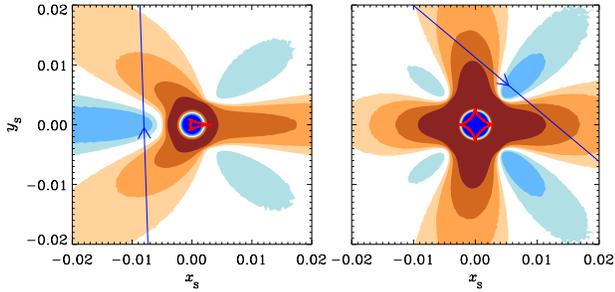}
\caption{\label{fig:one}
Central caustics induced by a planetary (left panel) and a binary companion 
(right panel). The regions with brownish and bluish colors represent the areas 
where the lensing magnification is higher and lower than the corresponding 
single-lensing magnification, respectively. For each tone, the color changes 
to darker shades when the fractional difference between the single and binary 
magnification is > 2$\%$, 4$\%$, 8$\%$, and 16$\%$, respectively. 
}\end{figure}

\begin{figure}[ht]
\epsscale{1.15}
\plotone{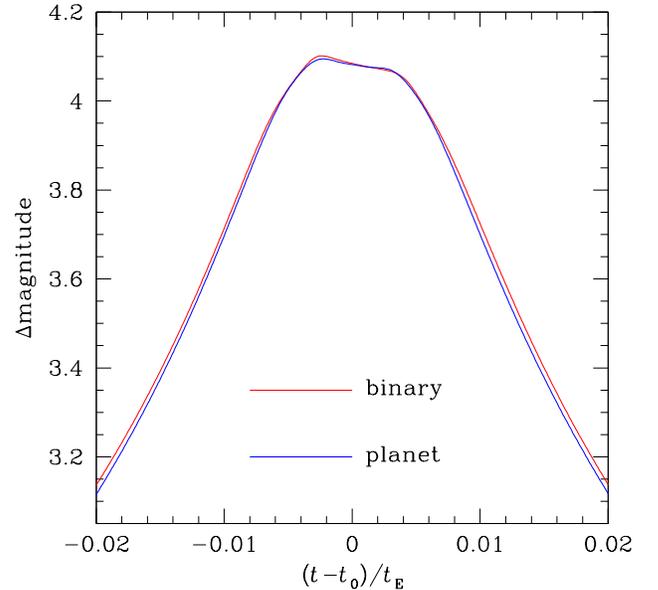}
\caption{\label{fig:two}
Light curves resulting from the two source trajectories (straight lines with arrows) 
marked in Fig. \ref{fig:one}. 
}\end{figure}

Perturbations near the peak of a high-magnification lensing event (central perturbations) 
can be produced not only by a planet but also by a binary companion \citep{han09,shin12}. 
For a binary lens where the projected separation between the lens components is 
substantially smaller than the Einstein radius (close binary), there exists a small 
single set of caustics formed around the barycenter of the binary. For a binary 
where the projected separation is substantially larger than the Einstein radius 
(wide binary), on the other hand, there exist two sets of caustics each of which 
is located close to each lens component. Then, for a high-magnification event 
resulting from the source trajectory passing close to the center of mass of a close 
binary or close to one of the lens components of a wide binary, there can be a short-term 
perturbation near the peak of the lensing light curve, similar to the central perturbation 
induced by a planet. It is known that the central perturbation induced by a planet can 
be generally distinguished from that induced by a binary because the caustic shapes and 
the resulting magnification patterns around the two types of caustics are different from each other.

In this paper, we present a case of central perturbations for which it is difficult 
to distinguish between the planetary and binary interpretations. In $\S$2, we describe 
details of the degeneracy. In $\S$3, we demonstrate the degeneracy for two microlensing 
events OGLE-2011-BLG-0526 and OGLE-2011-BLG-0950/MOA-2011-BLG-336 that were detected during 
the 2011 observation season. In $\S$4, we summarize the results and conclude.

\section{DEGENERACY}

\begin{deluxetable*}{ll}
\tablecaption{Telescopes\label{table:one}}
\tablewidth{0pt}
\tablehead{
\multicolumn{1}{c}{event} &
\multicolumn{1}{c}{telescopes}
}
\startdata
OGLE-2011-BLG-0526   & OGLE 1.3 m Warsaw telescope at Las Campanas Observatory in Chile                     \\
                     & MiNDSTEp 1.54 m Danish telescope at La Silla Paranal Observatory in Chile            \\
                     & PLANET 0.6 m at Perth Observatory in Australia                                       \\
                     & PLANET 1.0 m at SAAO in South Africa                                                 \\
                     & RoboNet 2.0 m Liverpool telescope (LT) in La Palma, Spain                            \\
\hline
OGLE-2011-BLG-0950/  & OGLE 1.3 m Warsaw telescope at Las Campanas Observatory in Chile                     \\
MOA-2011-BLG-336     & MOA 1.8 m at Mt. John Observatory in New Zealand                                     \\
                     & $\mu$FUN 1.3 m SMARTS telescope at CTIO in Chile                                     \\  
                     & $\mu$FUN 0.4 m at Auckland Observatory in New Zealand                                \\
                     & $\mu$FUN 0.4 m at Farm Cove Observatory (FCO) in New Zealand                         \\
                     & $\mu$FUN 0.4 m at Kumeu Observatory in New Zealand                                   \\
                     & $\mu$FUN 0.6 m at Observatorio do Pico Dos Dias (OPD) in Brazil                      \\
                     & $\mu$FUN 1.0 m at Wise Observatory in Israel                                         \\
                     & MiNDSTEp 1.54 m Danish telescope at La Silla Paranal Observatory in Chile            \\
                     & PLANET 1.0 m at SAAO in South Africa                                                 \\ 
                     & RoboNet 2.0 m Faulkes Telescope North (FTN) in Hawaii                                \\
                     & RoboNet 2.0 m Faulkes Telescope South (FTS) in Australia                             \\
                     & RoboNet 2.0 m LT in La Palma, Spain           
\enddata  
\end{deluxetable*}

The pattern of central perturbations in a lensing light curve is basically 
determined by the shape of the central caustic. For both planetary and binary 
cases, the central caustics form a closed figure that is composed of concave 
curves that meet at cusps. The general magnification pattern is that a positive 
perturbation occurs when the source is located in the region outside the caustic 
extending from cusps while a negative perturbation occurs when the source is 
located in the region between cusps. Here a ``positive'' (``negative'') perturbation 
means that the magnification of the perturbed part of the light curve is higher 
(lower) than the magnification of the corresponding single-lensing event. The central 
caustics induced by a planet and a binary companion have different shapes and thus 
the resulting patterns of magnification around the two types of caustics are 
different from each other. In Figure \ref{fig:one}, we present the central caustics and 
the magnification patterns around them for the representative cases of the planetary 
and binary lenses, respectively.

\begin{figure}[ht]
\epsscale{1.15}
\plotone{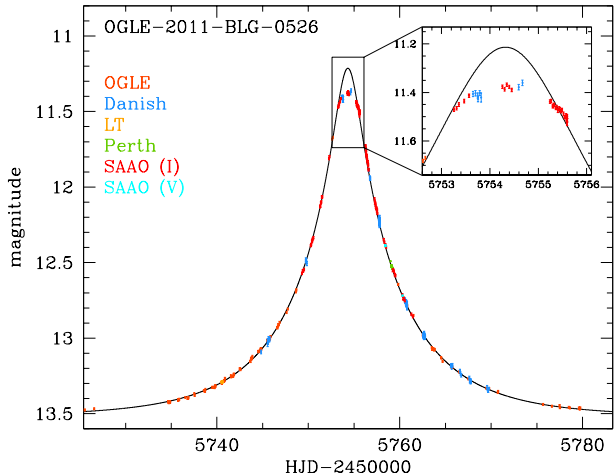}
\caption{\label{fig:three}
Light curve of OGLE-2011-BLG-0526. Also drawn is the best-fit single-lensing 
light curve that is obtained with data except those around the perturbation. 
Colors of data points are chosen to match those of the labels of observatories 
where the data were taken. The inset shows the enlarged view of the peak region.
}\end{figure}

\begin{figure}[ht]
\epsscale{1.15}
\plotone{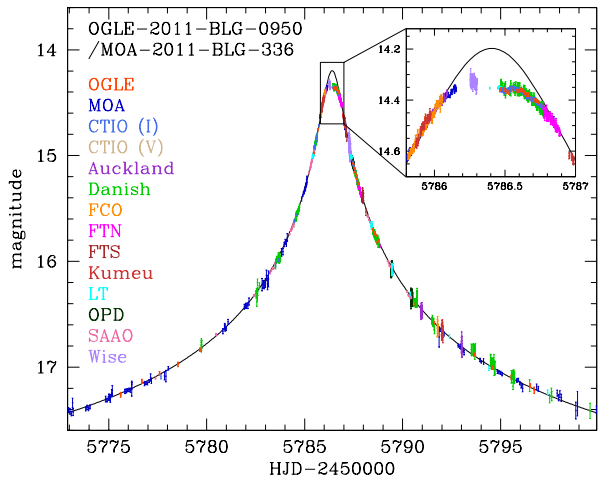}
\caption{\label{fig:four}
Light curve of OGLE-2011-BLG-0950/MOA-2011-BLG-336. Notations are same as those 
in Fig. \ref{fig:three}.
}\end{figure}

The central caustic induced by a planet has a shape of an arrowhead with four cusps. 
One cusp corresponding to the sharp tip of the arrowhead-shaped caustic is located 
on the star-planet axis. This cusp is strong in the sense that light curves resulting 
from source trajectories passing close to the cusp exhibit strong deviations from the 
single-lens expectation. Two other cusps are located off the star-planet axis corresponding 
to the blunt ends of the arrowhead-shaped caustic. These two cusps are moderately strong. 
The fourth cusp, which is located on the star-planet axis between the two off-axis cusps, 
is weak in the sense that it creates relatively weak deviations. Due to the weakness of 
the last cusp, there exists an extended region of negative perturbation between the two 
off-axis cusps.

The central caustic induced by a wide or a close binary has an asteroid shape with four 
cusps. Two of the cusps are located on the binary-lens axis and the other two are along 
a line perpendicular to the axis. The caustic is symmetric with respect to the two lines 
connecting the on-axis and off-axis cusps. Due to the symmetry of the caustic, 
all cusps are of similar strength. Regions of positive perturbations form outside the 
caustic extending from the cusps and regions of negative perturbations form between 
the positive-perturbation regions.

\begin{deluxetable*}{l|llll|llll}
\tablecaption{Best-fit Parameters\label{table:two}}
\tablewidth{0pt}
\tablehead{
\multicolumn{1}{c|}{parameter} &
\multicolumn{4}{c|}{OGLE-2011-BLG-0526} &
\multicolumn{4}{c}{OGLE-2011-BLG-0950/MOA-2011/BLG-336} \\
\multicolumn{1}{c|}{} &
\multicolumn{1}{c}{A} &
\multicolumn{1}{c}{B} &
\multicolumn{1}{c}{C} &
\multicolumn{1}{c|}{D} &
\multicolumn{1}{c}{A} &
\multicolumn{1}{c}{B} &
\multicolumn{1}{c}{C} &
\multicolumn{1}{c}{D} 
}
\startdata
$\chi^2$                   & 423.6              & 420.0  	                & 422.2                   & 422.9            & 3073.5                  & 2968.6                  & 2969.0                  & 3076.9                  \\
$u_0$                      & 0.141$\pm$0.001    & 0.117$\pm$0.002         & 0.117$\pm$0.002         & 0.140$\pm$0.020  & (9.3$\pm$0.1)$10^{-3}$  & (8.6$\pm$0.1)$10^{-3}$  & (8.7$\pm$0.1)$10^{-3}$  & (9.0$\pm$0.3)$10^{-3}$  \\
$t_{\rm E}$ (days)         & 11.63$\pm$0.08     & 12.15$\pm$0.09          & 12.37$\pm$0.10          & 11.60$\pm$1.91   & 61.39$\pm$0.67          & 65.21$\pm$0.85          & 65.27$\pm$0.76          & 62.41$\pm$1.90          \\
${\it s}$                  & 0.311$\pm$0.003    & 0.48$\pm$0.01           & 1.94$\pm$0.02           & 6.43$\pm$0.05    & 0.075$\pm$0.001         & 0.70$\pm$0.01           & 1.43$\pm$0.01           & 22.7$\pm$0.3            \\
${\it q}$                  & 0.91$\pm$0.04      & (3.5$\pm$0.2)$10^{-2}$  & (3.9$\pm$0.2)$10^{-2}$  & 28.5$\pm$10.6    & 0.83$\pm$0.09           & (5.8$\pm$0.2)$10^{-4}$  & (6.0$\pm$0.2)$10^{-4}$  & 2.36$\pm$0.21           \\
$\alpha$                   & -0.795$\pm$ 0.010  &	4.718$\pm$0.004         & 4.718$\pm$0.004         & 0.765$\pm$0.007  & 0.739$\pm$0.005         & 4.664$\pm$0.002         & 4.664$\pm$0.002         & 0.722$\pm$0.002         \\
$\rho_{\star}$ $(10^{-3})$ & 80$\pm$2           & --                      & --                      & 79$\pm$7         & 3.2$\pm$0.3             & 4.6$\pm$0.1             & 4.6$\pm$0.1             & 3.4$\pm$0.3             \\ 
$\pi_{{\rm E},N}$          & --                 & --                      & --                      & --               & 0.22$\pm$0.15           & -0.10$\pm$0.17          & -0.29$\pm$0.14          & 0.12$\pm$0.09           \\
$\pi_{{\rm E},E}$          & --                 & --                      & --                      & --               & -0.04$\pm$0.03          & 0.02$\pm$0.03           & 0.03$\pm$0.02           & -0.03$\pm$0.02            
\enddata  
\end{deluxetable*}

\begin{figure}[ht]
\epsscale{1.15}
\plotone{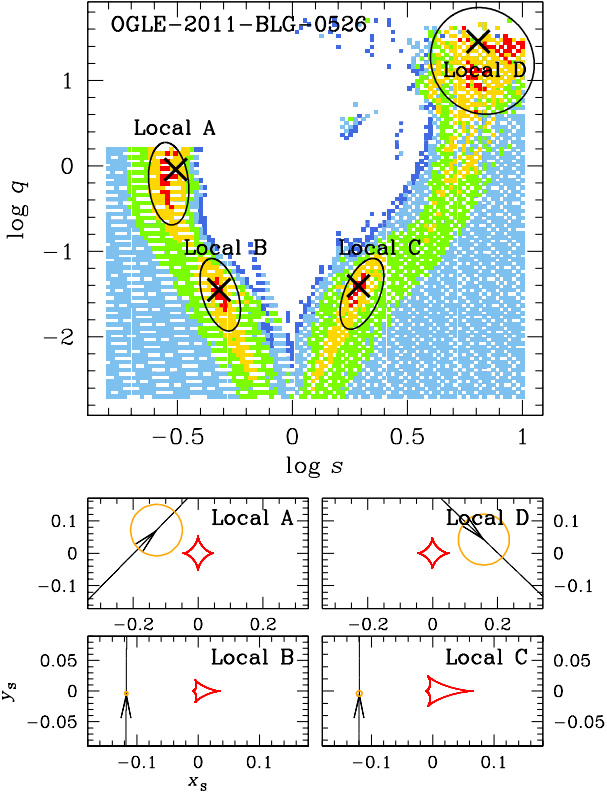}
\caption{\label{fig:five}
Distribution of $\Delta\chi^2$ in the parameter space of the projected binary separation 
($s$) and the mass ratio ($q$) for OGLE-2011-BLG-0526. The regions marked in red, yellow, 
green, sky blue, and blue correspond to those with $\Delta\chi^2 <$ $6^2$, $12^2$, $18^2$, $24^2$, and 
$30^2$, respectively. The cross marks represent the locations of the local minima. The lower 
panels show the source trajectories (straight lines with arrows) with respect to the caustics 
for the individual local solutions. The small orange circle on each source trajectory represents 
the relative scale of the source star. 
}\end{figure}

\begin{figure}[ht]
\epsscale{1.15}
\plotone{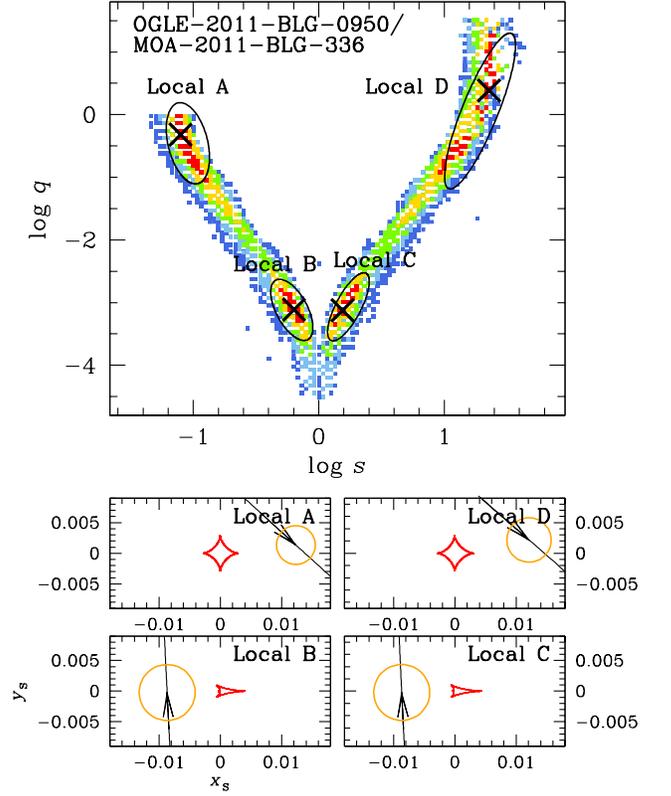}
\caption{\label{fig:six}
Distribution of $\Delta\chi^2$ in the $s-q$ parameter space for OGLE-2011-BLG-0950/MOA-2011-BLG-336. 
The regions marked in red, yellow, green, sky blue, and blue correspond to those with $\Delta\chi^2 <$ 
$13^2$, $26^2$, $39^2$, $52^2$, and $65^2$, respectively. Notations are same as in Fig. \ref{fig:five}.
}\end{figure}

Despite the basically different caustic shapes and the resulting magnification patterns, 
we find a case of central perturbations for which it is difficult to distinguish between 
the planetary and binary interpretations. This degeneracy is illustrated in Figures 
\ref{fig:one} and \ref{fig:two}. The planetary lensing case for this degeneracy occurs 
when the source trajectory passes the negative perturbation region behind the back end 
of the arrowhead-shaped central caustic with an angle between the source trajectory and 
the star-planet axis (source-trajectory angle) of $\alpha \sim 90^{\circ}$. For a binary 
case, a similar perturbation occurs when the source trajectory passes through the negative 
perturbation region between two cusps of an astroid-shaped caustic with a source-trajectory 
angle of $\sim 45^{\circ}$. For both cases, the morphology of the resulting perturbation 
is that the peak of the light curve appears to be blunt and flat.

\section{ACTUAL EVENTS}

We search for high-magnification events with similar central perturbations among 
those detected during the 2011 observation season. From this search, we find that two events 
including OGLE-2011-BLG-0526 and OGLE-2011-BLG-0950/MOA-2011-BLG-336 exhibit such central 
perturbations. In this section, we investigate the severity of the degeneracy by conducting 
detailed modeling of the light curves for these events.

\begin{figure}[ht]
\epsscale{1.15}
\plotone{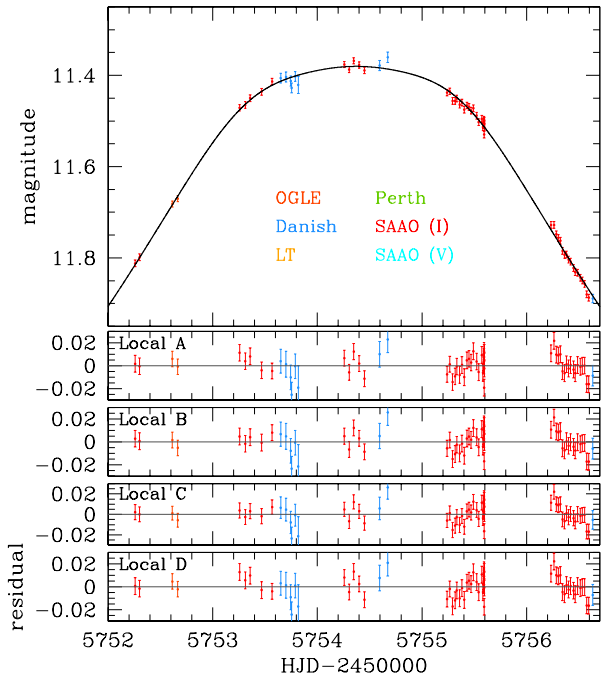}
\caption{\label{fig:seven}
Light curve of OGLE-2011-BLG-0526 near the peak region and the residuals from 4 local solutions. 
The model light curve drawn over the data is based on one of the local solutions (local ``B''). 
Colors of data points are chosen to match those of the labels of observatories where the data were taken.
}\end{figure}

The event OGLE-2011-BLG-0526 occurred on a Galactic bulge star that is positioned 
at $(\alpha,\delta)_{J2000}$ = $(18^{\rm h}02^{\rm m}45^{\rm s}\hskip-2pt.37,
-28^{\circ}01^{\prime}25^{\prime\prime}\hskip-2pt.8)$, 
which correspond to the Galactic coordinates $(l,b)$ = $(2.69^{\circ},-2.79^{\circ})$. 
The event was detected and alerted to the microlensing community by the Optical Gravitational 
Lensing Experiment (OGLE) group. High-magnification events are usually realerted after 
the first alert. Unfortunately, no high-magnification alert was issued for this event and 
thus follow-up observations were conducted by using a fraction of telescopes available for 
follow-up observations. As a result, the coverage of the peak is not very dense. The telescopes 
used for the observations of this event are listed in Table \ref{table:one}.

The event OGLE-2011-BLG-0950/MOA-2011-BLG-336 also occurred on a Galactic bulge star 
located at $(\alpha,\delta)_{J2000}$ = $(17^{\rm h}57^{\rm m}16^{\rm s}\hskip-2pt.63,
-32^{\circ}39^{\prime}57^{\prime\prime}\hskip-2pt.0)$, 
corresponding to $(l,b)$ = $(358.07^{\circ},-4.05^{\circ})$. It was independently 
discovered from the survey experiments conducted by the OGLE and the Microlensing Observation 
in Astrophysics (MOA) groups. A high-magnification alert was issued for this event 4 days before 
the peak. Based on this alert, follow-up observations were conducted by using 13 telescopes 
located in 8 different countries. As a result, the perturbation was more densely covered than 
the perturbation of OGLE-2011-BLG-0526. In Table \ref{table:one}, we also list the telescopes 
used for the observations of this event.

Initial reductions of the data taken from different observatories were processed by using photometry 
codes developed by the individual groups. For the purpose of improving the data quality, we conducted 
additional photometry for all follow-up data of OGLE-2011-BLG-0950/MOA-2011-BLG-336 by using 
codes based on difference imaging photometry. For the use of modeling, we rescaled the error bars 
of the data sets so that $\chi^{2}$ per degree of freedom becomes unity for each data set, 
where the value of $\chi^{2}$ is calculated based on the best-fit solution obtained from modeling. 
We eliminated 3$\sigma$ outliers from the best-fit solution in the modeling. 

\begin{figure}[ht]
\epsscale{1.15}
\plotone{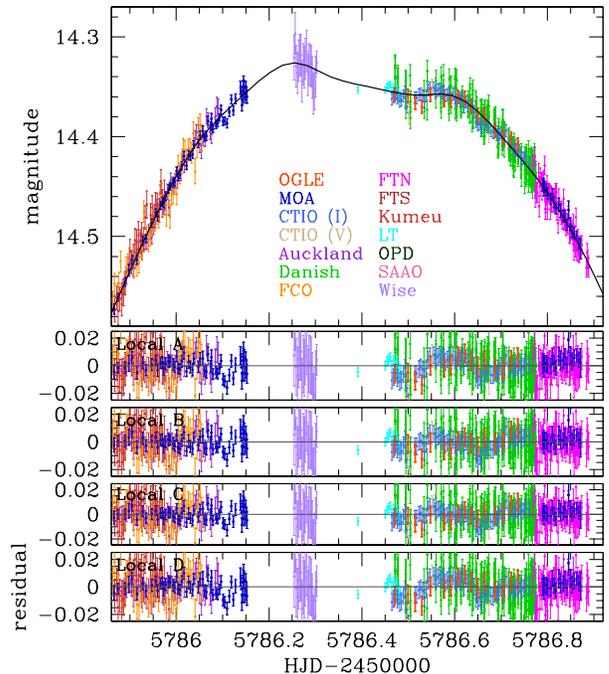}
\caption{\label{fig:eight}
Light curve of OGLE-2011-BLG-0950/MOA-2011-BLG-336 near the peak region and the residuals from 4 local 
solutions. The model light curve drawn over the data is based on one of the local solutions (local ``C''). 
Notations are same as those in Fig. \ref{fig:seven}.
}\end{figure}

In Figures \ref{fig:three} and \ref{fig:four}, we present the light curves of the two events. 
Also drawn are the best-fit single-lensing light curves. For both events, the light curves are 
well represented by those of standard single-lensing events except for the short-lasting perturbations 
near the peak. The common morphology of the perturbations is that the peak appears to be flat and blunt.

To investigate the nature of the perturbations, we conducted binary-lens modeling of the light curves. 
In the modeling of each light curve, we searched for the solution of the binary-lensing parameters that 
best describe the observed light curve by minimizing $\chi^{2}$ in the parameter space. For OGLE-2011-BLG-0526, 
the time scale of the event is not long ($t_{\rm E} \sim$ 12 days) and thus we modeled the light curve 
using 7 basic binary-lens parameters. The first 3 of these parameters characterize the geometry of 
the lens-source approach and they include the Einstein time scale, $t_{\rm E}$, the time of the closest 
lens-source approach, $t_0$, and the lens source separation at that moment, $u_0$, in units of the Einstein radius. 
The other 3 parameters characterize the binary lens. These parameters include the mass ratio between 
the lens components, $q$, the projected separation in units of the Einstein radius, $s$, and the angle between 
the source trajectory and the binary axis, $\alpha$. The last parameter of the normalized source radius 
$\rho_{\star}$ describes the deviation of the light curve affected by the finite-source effect and 
it represents the angular source radius $\theta_{\star}$ in units of the angular Einstein radius $\theta_{\rm E}$, i.e.\ 
$\rho_{\star}=\theta_{\star}/\theta_{\rm E}$. For OGLE-2011-BLG-0950/MOA-2011-BLG-336, the duration of the event 
($t_{\rm E} \sim$ 65 days) is relatively long. For such a case, the motion of the source with respect 
to the lens may deviate from a rectilinear one due to the change of the observer's position caused 
by the orbital motion of the Earth around the Sun and this deviation can cause a long-term deviation 
in the light curve \citep{gould92a}. Consideration of this ``parallax effect'' requires to include 
two additional parameters $\pi_{{\rm E},N}$ and $\pi_{{\rm E},E}$, which represent the two components 
of the lens parallax ${{\bf \pi}_{\rm E}}$ projected on the sky in the north and east equatorial coordinates, 
respectively. The direction of the parallax vector corresponds to the relative lens-source motion 
in the frame of the Earth at a specific time of the event. Its size corresponds to the ratio of 
the Earth's orbit to the physical Einstein radius, $r_{\rm E}$ = $D_{L}\theta_{\rm E}$, projected 
on the observer plane, i.e. $\pi_{\rm E}=({\rm AU}/r_{\rm E})[(D_{\rm S}-D_{\rm L})/D_{\rm S}]$.

Knowing that central perturbations can be produced either by a planet or by a binary companion, we conduct a 
thorough search for solutions in the $s-q$ parameter space encompassing both planet and binary regimes to 
investigate the possible existence of local minima. In Figures \ref{fig:five} and \ref{fig:six}, we present 
the resulting distributions of $\Delta\chi^2$ in the $s-q$ parameter space for the individual events. 
From the distributions, it is found that there exist four distinct local minima for both events. Among them, 
two minima are located in the region with $s > 1$ and the other two are located in the region with $s < 1$. 
For each close/wide binary pair, one local minimum is located in the regime of a binary mass ratio ($q \sim 1$) 
and the other minimum is located in the regime of a planet mass ratio ($q \ll 1$). We designate the individual 
minima by ``A'' ($s < 1$ with binary $q$), ``B'' ($s < 1$ with planetary $q$), ``C'' ($s > 1$ with planetary $q$), 
and ``D'' ($s > 1$ with binary $q$).

In Table \ref{table:two}, we present the lensing parameters of the individual local minima that are obtained 
by further refining the local solutions in the corresponding parameter space. The exact locations of the 
local minima are marked by ``X'' on the $\Delta\chi^2$ maps in Figures \ref{fig:five} and \ref{fig:six}. 
For each local solution, we also present the caustic and the source trajectory. We note that the size of 
the caustic for the binary with $s < 1$ is scaled by the Einstein radius corresponding to the total mass 
of the lens, while the caustic size for the binary with $s > 1$ is scaled by the Einstein radius corresponding 
to the mass of the lens component that the source approaches. 

The findings from the comparison of the local solutions and the corresponding lens-system geometries are summarized as below.

\begin{enumerate}
\item
For both events, $\chi^2$ differences from the best-fit single-lensing models are very big. We find that $\Delta\chi^2$ = 1085 
for OGLE-2011-BLG-0526 and $\Delta\chi^2$ = 5644 for OGLE-2011-BLG-0950/MOA-2011-BLG-336, implying that the perturbations 
of both events are clearly detected. 
\item
Despite the clear signature of the perturbation, we find that the degeneracy of the four local solutions is severe. 
To better show the subtle differences between the local solutions, we present the residuals of the data from the individual 
local solutions in Figures \ref{fig:seven} and \ref{fig:eight} for OGLE-2011-BLG-0526 and OGLE-2011-BLG-0950/MOA-2011-BLG-336, 
respectively. We also present the enlargement of the perturbed parts of the light curve in the upper panel of each figure. 
For the case of OGLE-2011-BLG-0526, the $\chi^2$ difference between the planetary and binary models is $\sim$ 3, implying 
that the degeneracy is very severe. For the case of OGLE-2011-BLG-0950/MOA-2011-BLG-336, the planetary solution is favored over 
the binary solution with $\Delta\chi^2 \sim$ 105 and thus the stellar binary model is formally excluded. However, from the visual 
inspection of the residuals, it is found that systematic residuals of the data from the planetary model are larger than 
the difference between the planetary and binary models. In addition, the CTIO, Danish, and OGLE data of overlapping 
coverage appear to be different from each other by an amount at least as large as the difference between the planetary and stellar 
binary models. Therefore, it is difficult to claim a planet discovery based on < 1\% variations in the light curve. 
\item
For a pair of solutions with similar mass ratios, the solutions with $s > 1$ and $s < 1$ result in a similar caustic shape. 
The degeneracy between these solutions, often referred to as $s \leftrightarrow s^{-1}$ degeneracy, is known to be caused by 
the symmetry of the lens-mapping equation between close and wide binaries \citep{dominik99,albrow99,afonso00,an05,chung05}.
\end{enumerate}

The degeneracy between the pairs of solutions with planetary and binary mass ratios corresponds to the degeneracy mentioned 
in $\S$ 2. To be noted is that despite the large difference in caustic shape, the resulting perturbations appear to be very 
alike. The planet/binary degeneracy introduced in this work was not known before. This is mostly because the caustics induced 
by a planet and a binary companion have very different shapes and thus it is widely believed that perturbations induced by 
the two types of companions can be easily distinguished. Considering that two events of a single season suffer from this 
degeneracy along with the fact that perturbations caused by non-caustic-crossing source trajectories have larger cross sections, 
it is expected that central perturbations suffering from this is common.

\section{CONCLUSION}

We introduced a new type of degeneracy in the planet/binary interpretation of central perturbations 
in microlensing light curves. The planetary lensing case for this degeneracy occurs when the source 
trajectory passes the negative perturbation region behind the back end of the arrowhead-shaped central 
caustic with a source-trajectory angle of $\sim 90^{\circ}$. For a binary case, a similar perturbation 
occurs when the source trajectory passes through the negative perturbation region between two cusps of 
an astroid-shaped caustic with a source-trajectory angle of $\sim 45^{\circ}$. For both cases, the 
morphology of the resulting perturbation is that the peak of the light curve appears to be blunt and flat. 
From investigation of events detected during the 2011 microlensing observation season, we found 2 events 
OGLE-2011-BLG-0526 and OGLE-2011-BLG-0950/MOA-2011-BLG-336, which exhibit such perturbations. From detailed 
modeling of the light curves, we demonstrated the severity of the degeneracy. Considering that 2 events 
during a single season suffer from the degeneracy, we conclude that central perturbations experiencing 
the degeneracy should be common.

\acknowledgments 
Work by CH was supported by Creative Research Initiative Program 
(2009-0081561) of National Research Foundation of Korea.
The OGLE project has received funding from the European Research Council 
under the European Community's Seventh Framework Programme 
(FP7/2007-2013) / ERC grant agreement no. 246678. The MOA experiment was 
supported by grants JSPS22403003 and JSPS23340064. 
TS was supported by the grant JSPS23340044.
Y. Muraki acknowledges support from JSPS grants JSPS23540339 and JSPS19340058.
The MiNDSTEp monitoring campaign is powered by ARTEMiS
(Automated Terrestrial Exoplanet Microlensing Search; Dominik et al.
2008, AN 329, 248). MH acknowledges support by the German Research
Foundation (DFG). DR (boursier FRIA), OW (FNRS research fellow) and J. Surdej acknowledge support
from the Communaut\'{e} fran\c{c}aise de Belgique Actions de recherche
concert\'{e}es -- Acad\'{e}mie universitaire Wallonie-Europe. KA, DMB,
MD, KH, MH, CL, CS, RAS, and YT are thankful to Qatar National Research
Fund (QNRF), member of Qatar Foundation, for support by grant NPRP
09-476-1-078.
CS received funding from the European Union Seventh Framework Programme (FPT/2007-2013) under grant agreement 268421.
This work is based in part on data collected by MiNDSTEp with the Danish 1.54 m telescope at the ESO La Silla Observatory. 
The Danish 1.54 m telescope is operated based on a grant from the Danish Natural Science Foundation (FNU). 
A. Gould and B.S. Gaudi acknowledge support from NSF AST-1103471.
B.S. Gaudi, A. Gould, and R.W. Pogge acknowledge support from NASA grant NNG04GL51G.
Work by J.C. Yee is supported by a National Science Foundation Graduate Research Fellowship under Grant No. 2009068160.
S. Dong's research was performed under contract with the California Institute of Technology (Caltech) funded by NASA through the Sagan Fellowship Program.
Research by TCH was carried out under the KRCF Young Scientist Research Fellowship Program. 
TCH and CUL acknowledge the support of Korea Astronomy and Space Science Institute (KASI) grant 2012-1-410-02.
Dr. David Warren provided financial support for Mt. Canopus Observatory.

\end{document}